\documentclass[reprint,superscriptaddress,amsmath,amssymb,
aps,pre,onecolumn,longbibliography,floatfix,]{revtex4-2}
\usepackage{graphicx}
\usepackage{euler,bm,enumerate}
\def \lapc {\nabla^2_{\rm sc}}
\def \ds {d{\bm s}}
\def \deltas {\delta{\bm s}}

\def \Ar {A_r}
\def \At {A_{\theta}}
\def \Ap {A_{\phi}}

\def \onebysth {\frac{1}{\sin\theta}}
\def \onebyr {\frac{1}{r}}
\def \onebyrsth {\frac{1}{r\sin\theta}}
\def \Aij {\frac{\partial A_i}{\partial x^j}}
\def \ddtheta {\frac{\partial}{\partial\theta}}
\def \ddphi {\frac{\partial}{\partial\phi}}
\def \ddphisqr {\frac{\partial^2}{\partial\phi^2}}
\def \ddr {\frac{\partial}{\partial r}}
\def \ll {\hat{\lambda}}
\def \mm {\hat{\mu}}
\def \AA {{\bm A}}

\def \Br {B_r}
\def \Bt {B_{\theta}}
\def \Bp {B_{\phi}}

\def \grad {{\bm \nabla}}
\def \curl {{\bm \nabla} \times}
\def \dive {{\bm \nabla}\cdot}
\def \lap {\nabla^2}

{}

\def \BB {\bm B}

\def \Ar {A_r}
\def \curl {{\bm \nabla}\times}

\def \BB {\bm B}

\newcommand{\eq}[1]{(\ref{#1})}

\newcommand{\deldel}[2]{\frac{\partial #2}{\partial x_{#1}}}
\begin{document}

\title{Vector Differential Operators in arbitrary coordinates : 
a general approach}
\author{Priyabrata Mitra}
\email{priyabrata.mitra@gmail.com}
\affiliation{BL 76, Sector II, Salt Lake, Kolkata 700091}
\author{Dhrubaditya Mitra}%
\email{dhrubaditya.mitra@su.se}
\affiliation{Nordita, KTH Royal Institute of Technology and
Stockholm University, Hannes Alfv\'ens v\"ag 12, 10691 Stockholm, Sweden}

\begin{abstract}
  We present a method for calculating the results of operation of differential
  operators operating on components of vector in generalized coordinates not
  restricted to orthogonal one.
  For this we use the relationships between covariant, contravariant and
  physical components of a vector and the idea of covariant differentiation.
  This not only simplifies vector calculus in common curvilinear coordinates,
  e.g., cylindrical or spherical polar, but also provides a deeper
  understanding of these operators in coordinate independent form.  
\end{abstract}

\maketitle

\section{Introduction}
Undergraduate students in physics and engineering need vector differential
operators in curvilinear coordinates quite early in their studies,
for example, in courses of  electricity and magnetism or
elasticity and fluid dynamics.
It is a straightforward but tedious exercise to obtain
the expressions of these operators from their Cartesian
expressions by direct substitution.
Most textbooks also provide the expressions of some of these operators
for commonly used coordinate systems, e.g., cylindrical and
spherical polar.
In contrast, here we present a method based on elementary tensor
analysis to obtain the expressions of vector differential operators
in any coordinate systems, even non-orthogonal ones.
Here we present such a general method that is based on elementary
tensor analysis. 
As our treatment is aimed at the beginners, we do not use the formal
language of modern differential geometry.
We only assume familiarity with calculus of several variables.

To be specific, we use two concepts from standard tensor analysis: (a)
the concept of contravariant, covariant, and physical component of a
tensor and (b) the covariant differentiation.
We provide 
a detailed description of the first and a brief description
of the second which is also 
covered in many elementary textbooks, e.g., Ref.~\cite{Aris62}.

It is well known that~\cite{SomII} the expression for $\lap \AA$ in 
curvilinear coordinates calculated from the  vector identity
\begin{equation}
\lap\AA = \grad\dive \AA - \curl\curl \AA
\end{equation} 
does not agree with the expression obtained
by application of the scalar Laplacian operator on each of the components of
$\AA$, $A_i$,  in that coordinate system.
In fact the agreement is not expected, since the components of the vector,
$A_i$, are not  invariant scalars.
The usual expression for the Laplacian operator in curvilinear coordinates
assume the scalar nature of the operand. 
The Laplacian operator itself is essentially a scalar one. 
The result is a scalar when it acts on scalar functions. 
Operating on a vector function it yields another vector  function.
In order to find the proper expression that would result when 
Laplacian operator operates on a  component of a vector in any
curvilinear coordinates we are to perform covariant differentiation 
in place of partial differentiation.

Covariant differentiation operates on covariant/contravariant components of
a vector field and not on the physical components of a vector. 
Physical components at a point along the coordinate lines
are scalars, obtained by performing scalar 
product of the vector with unit vectors along that direction
at that point.
They are in general neither the covariant nor 
the contravariant component of a vector.
The second order covariant differentiation 
operating on a covariant component of a 
vector produces, in general, a covariant tensor of rank three. 
A physical component of this entity is obtained by 
contracting it with unit vector along a coordinate direction. 
The process of contraction in tensor corresponds to the scalar product
in vector algebra. 

Here we develop a general method along the lines described above to
calculate any vector differential operator in any coordinate system
(for which a metric can be defined). 
The method developed here is valid for any coordinate system including 
non-orthogonal ones. 

\section{The contravariant, covariant and physical
components of a vector}

The square of the distance between two neighboring points with
coordinates $x^i$ and $x^i+dx^i$ where $x^i$s are arbitrary
coordinates, is given by the line element
\begin{equation}
ds^2=g_{ij}dx^idx^j
\label{metric}
\end{equation}
where $g_{ij}$ are the components of metric tensor 
of the chosen coordinate system. From equation~\ref{metric} we get
\begin{equation}
1=g_{ij}\lambda^i\lambda^j
\label{lambda}
\end{equation}
where $\lambda^i = dx^i/ds$ are called parameters of the 
direction of the infinitesimal displacement $\ds$.

The parameters can be lowered by the metric tensor; i.e., 
\begin{equation}
\lambda_i=g_{ij} \lambda^j \/, 
\end{equation}
where $\lambda_i$ are called the moments of the direction.
For a given direction $\lambda^i(\lambda_i)$ are contravariant
(covariant) components of unit vector in that direction. 

Thus a vector $\AA$ can be represented by its physical components or by its
magnitude and direction. The latter being determined by its
parameters or moments. In that case, it can be represented by the quantities
\begin{subequations}
  \begin{align}
A^i &= A\lambda^i
\label{Acon} \\
\text{or}\quad
A_i&= A\lambda_i \/.
\label{Aco}
  \end{align}
\end{subequations}
Here $A^i$($A_i$) are called the contravariant (covariant)
components of the vector $\AA$. 

Let $\ll$ and $\mm$ be the unit vectors along two infinitesimal 
displacements $\ds$ and $\deltas$ from a generic point $P$ 
and $\psi$ be the angle between them. It can be shown~\cite{Lev26}
\begin{equation}
\cos \psi = g_{ij}\lambda^i\mu^j = \lambda^i\mu_i = \lambda_j\mu^j
\label{psi}
\end{equation}
For an infinitesimal displacement along the $x_i$ coordinate
\begin{equation}
ds^2=g_{ii} (dx^i)^2, \quad (i\quad\text{not summed})
\end{equation}
and $dx^j =dx^k = 0$. Using the relationship \eq{lambda} we
get the expression of the parameter for $x^i$-coordinate line
\begin{equation}
\lambda^i= \frac{1}{\sqrt{g_{ii}}}
\label{lami}
\end{equation}

From equation \ref{psi} we can find explicit relation between the physical
components and the covariant components of a vector. 
Suppose a vector $\AA$, with its parameters $\mu^j$, makes an angle $\psi$ 
with the $x_i$ coordinate line.
By definition, the physical component, $A_{x_i}$ is given by
 (\emph{for a fixed value of $i$})
\begin{equation}
A_{x_i} \equiv A\cos\psi = A g_{ij} \lambda^i \mu^j = g_{ij} \lambda^i A^j = \lambda^i A_i
\label{eq:Aphy}
\end{equation}
Here,  the equalities follow from equation \ref{psi} and the definition of 
contravariant (\ref{Acon}) and covariant (\ref{Aco}) components of a vector. 
Hence, the physical component and the covariant component along $x_i$ 
are related by 
\begin{equation}
A_{x_i} = \frac{A_i}{\sqrt{g_{ii} } }  \hspace{1cm} (i\; {\rm not\; summed} )
\end{equation}
Apart from a scale factor the covariant component is the orthogonal projection of $\AA$
along a coordinate line. 

\subsection{Two dimensional oblique coordinate system.}
The distinction between the contravariant, the covariant, and the physical components is most
clearly seen in non-orthogonal coordinates. We illustrate this by choosing the
two dimensional oblique coordinate system with an angle $\alpha$ ($\neq \pi/2$) 
between the coordinate lines $x_1$ and $x_2$  
The line element in this case is 
\begin{equation}
ds^2 = (dx^1)^2+2 \cos\alpha (dx^1)(dx^2)+(dx^2)^2
\label{ds2_ob}
\end{equation}
The coordinates of a point in this plane is obtained by drawing lines parallel to the 
axes from this point, i.e., by parallel projection, on the axes. 
By definition, the contravariant components of a vector transforms like coordinate differentials. 
Thus if the initial point of a vector $\AA$ is chosen as the origin of the coordinate
system then its contravariant components are identical to the coordinates of the tip of the vector. 

For the line element given in equation \ref{ds2_ob} 
\begin{subequations}
\begin{align}
g_{ij} &= \left( \begin{matrix} 1 & \cos\alpha \\ 
                                     \cos\alpha & 1 \end{matrix} \right ),\\
g^{ij} &= \frac{1}{\sin^2\alpha}\left( \begin{matrix} 1 & -\cos\alpha \\ 
  -\cos\alpha & 1 \end{matrix} \right )\\
\text{and}\quad |g_{ij}| &= \sin^2\alpha
\end{align}
\label{gij_ob}
\end{subequations} 
The physical components of $\AA$ along the two coordinate lines are,
(using equation \ref{psi} )
\begin{eqnarray}
A_{x_1} &=& A_1 = A^1 + A^2\cos\alpha, \\
A_{x_2} &=& A_2 =A^1\cos\alpha + A^2
\label{as}
\end{eqnarray} 
 respectively.  Solving for $A^1$ and $A^2$ simultaneously we obtain
\begin{eqnarray}
A^1 &=& \frac{1}{\sin^2\alpha} \left( A_{x_1} -A_{x_2} \cos\alpha \right), \\             
A^2 &=& \frac{1}{\sin^2\alpha} \left(- A_{x_1} \cos\alpha-A_{x_2}\right), 
\label{a12}
\end{eqnarray}
 We identify that in this coordinate system, the covariant components 
are orthogonal projection and contravariant components are parallel
projections of the vector on the coordinate lines. 
In orthogonal Cartesian coordinates, the orthogonal and parallel 
projections coincide, furthermore $\alpha = \pi/2$,  hence the covariant and 
contravariant components  are one and the same and identical with the physical 
components.  Because of these, the expression for $\lap \AA$ presents no difficulty in 
Cartesian coordinates. 

\subsection{Curvilinear Coordinates}

In curvilinear coordinates covariant and contravariant components differ from 
each other and from the physical components even when the coordinate system
is orthogonal. For spherical polar coordinates (SPC) we have 

\begin{equation}
ds^2 = dr^2 + r^2 d\theta^2 + r^2 \sin^2\theta d\phi^2 \/.
\end{equation}
For this line element 
\begin{subequations}
\begin{align}
 \lambda^1 &=1, \qquad  &\lambda^2 &= \frac{1}{r}, \qquad & 
          \lambda^3 &= \frac{1}{r\sin\theta}\/;
                                   \\
 \lambda_1 &=1, \qquad &\lambda_2 &= r,  \qquad &\lambda_3 &= r\sin\theta \/.
\label{eq:lambdasph}
\end{align}
\end{subequations}
Let us denote the three physical components by $\Ar$, $\At$, and
$\Ap$. 
Then,
\begin{subequations}
\begin{align}
\Ar &= \lambda^1 A_1 = A_1 = A^1 
\label{ar} \\
\At &= \lambda^2 A_2 = \frac{1}{r}A_2 =r A^2 
\label{at} \\
\Ap &= \lambda^3 A_3 = \frac{1}{r\sin\theta}A_3 = A^3 r\sin\theta 
\label{ap}
\end{align}
\end{subequations}
Inverting, we have the expression for the following covariant
and contravariant components of the vector $\AA$ in terms
of its physical components,
\begin{subequations}
\begin{align}
A_1 &= A_r \/, \quad  &A^1 = A_r \/,\\
A_2 &= r\At \/, \quad &A^2 = \frac{\At}{r} \/,\\
A_3 &= r\sin\theta \Ap\/, \quad  &A^3 = \frac{\Ap}{r\sin\theta} \/. 
\end{align}
\label{eq:asph}
\end{subequations}
\subsection{Relation between the physical components and 
covariant (contravariant) components in case of 
tensors of higher rank}

It is well known that in curved space or in flat space with curvilinear 
coordinates ordinary derivatives are to be replaced by covariant derivatives.
After covariant differentiation we get a tensor of next 
higher rank. Therefore to compare the results with those of standard
vector analysis we need a relationship between covariant
components and its physical one for tensor of higher rank. 
To obtain this we consider the direct product of two vectors
which constitutes a tensor of rank two. 
For example let $\AA$ and $\BB$ be two different vector fields. 
Using \eq{eq:Aphy} we obtain
\begin{equation}
A_{x_i}B_{x_j} = \lambda^i\lambda^j A_iB_j \/.
\end{equation}
Thus in general the physical components of a 
covariant tensor of rank two will be obtained
by contracting its components by proper parameters. 
That is
\begin{equation}
T_{x_ix_j} = \lambda^i \lambda^j T_{ij}
\label{eq:physical}
\end{equation}
where $T_{ij}$ are covariant components and $T_{x_ix_j}$ are 
corresponding physical components of the tensor field.

\section{Vector differential operators}

Let us now state again the main point of this paper. To calculate
components of vector differential operators operating on vector fields
along coordinate directions we do the following:
\begin{enumerate}[(a)]
\item Replace the partial derivatives by covariant derivatives and
operate them on the covariant component of the vector field to 
construct a tensor of higher rank. 
\item Calculate the physical component of this higher rank tensor along
a coordinate direction following the prescription of \eq{eq:physical}.
\item Substitute the covariant components of the vector field by their
physical components. 
\end{enumerate}
\subsection{Curl of a vector field}
We illustrate this procedure by calculating the components of curl of a
vector field in spherical polar coordinates. 
The covariant derivative of $A_i$ with respect to arbitrary coordinates
($x_j$) is given by~\cite{Lev26}
\begin{equation}
A_{i;j}=\frac{\partial A_i}{\partial x^j} - \Gamma_{ij}^kA_k
\label{eq:coderiv}
\end{equation}
We denote the covariant derivative with respect to a coordinate $x_j$ by 
putting a semi-color(;) before the suffix. The $\Gamma_{ij}^k$
are Christoffel's symbols of second kind defined by 
\begin{equation}
\Gamma_{ij}^m = g^{mk}\Gamma_{ij,k} = 
     \frac{1}{2}g^{mk} \left( 
\deldel{j}{g_{ik}}+\deldel{i}{g_{jk}}-\deldel{k}{g_{ij}}
\right)
\label{chris}
\end{equation}
From \eq{eq:coderiv} we get
\begin{equation}
A_{i;j}-A_{j;i} = \deldel{j}{A_i} - \deldel{i}{A_j}
\label{curl}
\end{equation}
Thus curl of a covariant component of tensor of rank one reduced to 
ordinary curl.
In general, the Contraviant component does not yield to such simplification.
As an example consider the spherical polar coordinate system. 
The non-vanishing Christoffel's symbol of the second kind in spherical 
polar coordinate are 
\begin{subequations}
\begin{align}
\Gamma_{22}^1& = -r \/, \quad  &\Gamma_{33}^1 &= -r \sin^2\theta,  \nonumber \\
\Gamma_{12}^2 &= \Gamma_{21}^2 = \frac{1}{r} \/, \quad & \Gamma_{33}^2 &= -\sin\theta\cos\theta,  \nonumber \\
\Gamma_{13}^3 &= \Gamma_{31}^3= \frac{1}{r} \/,\quad  & \Gamma_{23}^3 &= \Gamma_{32}^3 = \cot\theta \/. 
\label{eq:chrissph}
\end{align}
\end{subequations}
Following the prescription above let us first construct the tensor $G$
with covariant components $G_{ij}$ such that 
\begin{equation}
G_{ij} \equiv A_{j;i}
\end{equation}
The physical component of $G$ is given by 
\begin{equation}
G_{x_ix_j} = \lambda^i \lambda^j G_{ij}
\end{equation}
and the physical component of  $\BB = \curl \AA$ is obtained by
anti-symmetrizing the physical
components of $G$, i.e.,
\begin{equation}
B_{x_i} = \epsilon_{ijk} G_{x_jx_k}
\end{equation}
In particular,
\begin{eqnarray}
B_r &=& \lambda^2\lambda^3\left[A_{3;2} - A_{2;3} \right] \nonumber \\
&=&  \frac{1}{r^2\sin\theta}\left[ \ddtheta(A_3) - \ddphi(A_2)
                               \right]   \nonumber \\
&=& \frac{1}{r\sin\theta} \left[\ddtheta(\sin\theta\Ap) -
    \ddphi(\At) \right] \/.
\label{eq:curl23}
\end{eqnarray}
Where the second step follows from the symmetries of the Christoffel
coefficients and the final step follows from \eq{at} and \eq{ap}.
Similarly,
\begin{eqnarray}
B_{\theta} &=& \lambda^3\lambda^1(A_{1;3}-A_{3;1}) \nonumber \\
              &=& \frac{1}{r\sin\theta} 
                 \ddphi A_r - \frac{1}{r} \ddr \left(r\Ar \right)
\label{eq:curl31}
\end{eqnarray}
and
\begin{eqnarray}
B_{\phi} &=& \lambda^1\lambda^2(A_{2;1}-A_{1;2}) \nonumber \\
              &=& \frac{1}{r} \left( 
                 \ddphi (r\At) - \ddtheta\Ar \right)
\label{eq:curl12}
\end{eqnarray}
If instead of anti-symmetrizing we would have set $x_i = x_j$ and
summed, we would have obtained the expression for divergence of $\AA$
in spherical polar coordinate system. 
\subsection{Laplacian of a vector field}
As a second example, let us now proceed to calculate the Laplacian of a
vector field in a similar manner. 
\begin{enumerate}[(a)]
\item  Define covariant component of $\Lambda$, which is a tensor
of rank 3,  to be 
$\Lambda_{ijk} = A_{i;jk}$,
by calculating two covariant derivatives on $\AA$.  
\item Obtain the physical component 
$\Lambda_{x_ix_jx_k} = \lambda^i\lambda^j\lambda^kA_{i;jk}$
\item Set $x_j=x_k$ and sum to obtain the physical component of $\lap
  \AA$ along the  $x_i$ direction. 
\end{enumerate}
Let us remind our reader here that the 
the general expression for the second order covariant differentiation on a
covariant component of a vector field $\AA$ in a generalized coordinate
$(x^i)$ is given by 
\begin{widetext}
\begin{equation}
A_{i;jk} = \frac{\partial}{\partial x^k} 
             \left( \Aij - \Gamma^{m}_{ij}A_{m} \right)
     -\Gamma_{ki}^{n}\left(\frac{\partial A_{n}}{\partial x^j}
                - \Gamma^{m}_{\beta j} A_{m} \right )
     -\Gamma_{kj}^{n}\left(\frac{\partial A_{i}}{\partial x^{n}}
                - \Gamma^{m}_{in} A_{m} \right )
\label{eq:covar2}
\end{equation}
We illustrate this by calculating the Laplacian in spherical polar
coordinate system. Using the Eqns (\ref{eq:covar2}) and (\ref{eq:chrissph})
we get 
\begin{subequations}
  \begin{align}
A_{1;11} &= \ddr\left(\frac{\partial \Ar}{\partial r} \right ) , 
\label{eq:a111}\\
A_{1;22} &= \ddtheta\left( \frac{\partial \Ar}{\partial \theta} -\At \right ) 
           -\onebyr \left[ \ddtheta(r\At) + r\Ar \right] + 
           r\left( \frac{\partial A_r}{\partial r} \right ) ,
\label{eq:a122} \\
A_{1;33} &= \ddphi\left( \frac{\partial A_r}{\partial \phi} - 
                                      \sin\theta A_{\phi}\right )
             -\onebyr\left[ \ddphi(r\sin\theta\Ap) + r\sin^2\theta\Ar
                           + r\sin\theta\cos\theta\At \right]  \nonumber \\
         &    +r\sin^2\theta\left(\frac{\partial \Ar}{\partial r} \right ) +
               \sin\theta\cos\theta \left[\frac{\partial \Ar}{\partial \theta} 
               -\At \right]
\label{eq:a133}
\end{align}
\end{subequations}
\end{widetext}
where we have expressed the covariant components in terms of physical
components using Eqns(\ref{eq:asph}). Using Eqns~(\ref{eq:a111}),(\ref{eq:a122})
and (\ref{eq:a133}) we get
\begin{subequations}
  \begin{align}
\lap\Ar &= (\lambda^1)^3 A_{1;11} + \lambda^1(\lambda^2)^2 A_{1;22} + 
         \lambda_1(\lambda_3)^2 A_{1;33} \nonumber \\
&= \lapc \Ar - \frac{2}{r^2} \left[\Ar+\onebysth\ddtheta(\sin\theta\At) 
       + \onebysth\ddphi\Ap \right] \\
\lap\At  &= \lapc \At 
          - \frac{2}{r^2}\left[\ddtheta\Ar -\frac{\At}{2\sin^2\theta}
                               -\onebysth\cot\theta\ddphi\Ap \right] \\ 
\lap\Ap  &= \lapc \Ap 
          - \frac{2}{r^2\sin^2\theta}\left[\ddphi\Ap 
                                            +\cot\theta\ddphi\At
                                           -\frac{\Ap}{2\sin\theta} \right] \\
\text{where}\quad
\lapc &\equiv \frac{1}{r^2}\ddr\left(r^2\ddr\right) + 
\frac{1}{r^2\sin\theta} \ddtheta\left(\sin\theta\ddtheta \right)
+ \frac{1}{r^2\sin^2\theta}\ddphisqr
\label{eq:lap_scalar}
\end{align}
\end{subequations}
  is the usual Laplacian operator in spherical polar coordinate for scalar
operands. 

\subsection{ $(\AA\cdot\grad)\BB$ in spherical polar coordinate}
Following the same prescription we should proceed as follows:
\begin{enumerate}[(a)]
\item Construct $\Lambda_{ijk} = A_iB_{j;k}$
\item The physical component $\Lambda_{x_ix_jx_k} =
  \lambda^i\lambda^j\lambda^k\Lambda_{ijk}$
\item Set $i=k$ and sum to obtain the physical component along $x_j$. 
\end{enumerate}
In spherical polar coordinate the expression are 
\begin{subequations}
  \begin{align}
(\AA\cdot\grad)\Br &= (\AA\cdot\grad)_{\rm sc} \Br 
              - \frac{1}{r}\left(\At\Bt + \Ap\Bp \right), \\
(\AA\cdot\grad)\Bt &= (\AA\cdot\grad)_{\rm sc} \Bt 
              - \frac{1}{r}\left(\At\Br -\cot\theta \Ap\Bp \right), \\
(\AA\cdot\grad)\Bp &= (\AA\cdot\grad)_{\rm sc} \Bp 
              - \frac{1}{r}\left(\Ap\Br +\cot\theta \Ap\Bt \right)\\
\text{where}\quad
(\AA\cdot\grad)_{\rm sc} &= A_r\ddr + \At\onebyr\ddtheta+\Ap\onebyrsth\ddphi
  \end{align}
\end{subequations}
is the form of the operator operating on a scalar operand.

Expressions for few additional operators are given in the appendix. 
\section{Conclusion}
In summary, we present methods to calculate vector differential operators
in any generalized coordinate system whose metric is known.
Our method uses two concepts of tensor analysis, the definition of contravariant
covariant, and physical components of a tensor and the definition of
covariant differentiation.
We provide a description of the first.
We enthusiastically recommend the excellent book~\citep{Aris62} by
Aris for the second. 
Our method works even for non-orthogonal coordinate systems.

In our experience, the complex expression of vector differential
operators in curvilinear coordinates intimidates and confuses the students.
Our goal is to show how those expressions arise naturally
and inevitably as a consequence of general tensor analysis which
is truly the appropriate tool to use in generalized coordinates.
Generalized coordinates system are now becoming increasingly used
not only by physics students studying general relativity but also
by engineering and biology students interested in two-dimensional
elasticity.
We hope this presentation will be useful to them.

We intentionally avoid a formal representation such that anyone with
a knowledge of calculus of several variables can follow.
Our philosophy is best summarized by a quote from the
famous book ``Calculus Made Easy'' by Silvanus P. Thompson

``You don't forbid the use of a watch to every person who does not know how
to make one? You don't object to the musician playing on a violin that he has
not himself constructed. You don't teach the rules of syntax to children
until they have already become fluent in the use of speech.
It would be equally absurd to require general rigid demonstrations to be
expounded to beginners in the calculus.''


\appendix
\section{The expression for $\grad(\AA\cdot\BB)$ in spherical polar coordinate}
The $i$-th component of $\grad(\AA\cdot\BB)$ in an arbitrary curvilinear 
coordinate is 
\begin{equation}
\nabla_{x^i}(\AA\cdot\BB) = \lambda^i(A^lB_l)_{;i} \quad(i\; \text{held fixed})
\end{equation}
From the properties of covariant derivatives, Eq.~(\ref{eq:lambdasph}), we have
\begin{equation}
(A^lB_l)_{;i} = A^lB_{l;i} + A_{l;i} B_l \/.
\label{eq:abi}
\end{equation} 
Using Eqns (\ref{eq:coderiv}), (\ref{eq:curl23}), (\ref{eq:curl31}), 
(\ref{eq:curl12}), (\ref{eq:chrissph}), and (\ref{eq:asph}) we get for
spherical polar coordinates
\begin{equation}
A^lB_{l;r} = (\AA\cdot\grad)\Br + [\AA\times(\curl\BB)]_r - 
             \onebyr\left(\At\Bt+\Ap\Bp\right) \/.
\end{equation}
Interchanging $\AA$ and $\BB$ we the expression for the second term in 
Eq.~(\ref{eq:abi}). After contracting with $\lambda^1$ we get 
\begin{equation}
[\grad(\AA\cdot\BB)]_r = (\AA\cdot\grad)\Br + (\BB\cdot\grad)\Ar
                        +[\AA\times(\curl\BB)]_r + [\BB\times(\curl\AA)]_r
                      -\frac{2}{r}\left(\At\Bt + \Ap\Bp \right)
\end{equation}
Similarly we get for the other components
\begin{eqnarray}
[\grad(\AA\cdot\BB)]_{\phi} &=& (\AA\cdot\grad)\Bp + (\BB\cdot\grad)\Ap
              +[\AA\times(\curl\BB)]_{\phi} 
                           + [\BB\times(\curl\AA)]_{\phi} \nonumber \\
             &&+  \onebyr\left(\Ap\Br+\Ar\Bp+2\cot\theta\At\Bp \right)
\end{eqnarray}

\section{Expression for the strain tensor in curvilinear coordinate}
The symmetric strain tensor in Cartesian coordinates $(y_i)$ is defined by 
\begin{equation}
S_{ik} = \frac{1}{2}\left(\frac{\partial A_i}{\partial y_k} +  
                       \frac{\partial A_k}{\partial y_i} \right )
\label{eq:strain}
\end{equation}
where the $A_i$s now denote Cartesian components of the displacement of an
arbitrary point due to small deformation. In arbitrary curvilinear 
coordinates ($x^i$)  Eq.~(\ref{eq:strain}) reads
\begin{equation}
S_{x^ix^k} = \frac{1}{2}\lambda^i\lambda^k
                    \left(A_{i;k} +  A_{k;i} \right ) \/.
\label{eq:arbstrain}
\end{equation}
Following are explicit expressions of the components of this
tensor in spherical polar coordinate
\begin{eqnarray}
S_{rr} &=&  (\lambda^1)^2 A_{1;1} = \ddr \Ar \/, \\
S_{\theta\theta} &=&  (\lambda^2)^2 A_{2;2} 
         = \onebyr\left(\ddtheta \At + \Ar \right ) \/, \\
S_{\phi\phi} &=&  (\lambda^3)^2 A_{3;3} 
         = \onebyrsth\left(\ddphi \Ap + \cos\theta\At+\sin\theta\Ar \right) \/, \\
S_{\theta,\phi} &=& \frac{1}{2}\lambda^2\lambda^3\left( A_{2;3}+ A_{3;2} \right) 
                  \nonumber \\
          &=& \frac{1}{2}\onebyr\left(\onebysth\ddphi\At + \ddtheta\Ap - 
                      \cot\theta\Ap \right ) \\
S_{\phi r} &=& \frac{1}{2}\lambda^3\lambda^1\left( A_{3;1}+ A_{1;3} \right)
                           \nonumber \\
          &=& \frac{1}{2}\left(\ddr\Ap + \onebyrsth\ddphi\Ar - 
                      \onebyr\Ap \right ) \\
S_{r\theta} &=& \frac{1}{2}\lambda^1\lambda^2\left( A_{1;2}+ A_{2;1} \right) 
                        \nonumber\\
          &=& \frac{1}{2}\left(\ddtheta\Ar + \ddr\At - 
                      \onebyr\At \right ) 
\end{eqnarray}
\end{document}